\documentclass[journal]{IEEEtran}
\ifCLASSINFOpdf
\else
\fi

\usepackage[sorting=none]{biblatex}
\usepackage{amsmath,graphicx}

\usepackage{adjustbox}
\usepackage{fancyhdr}

\fancypagestyle{head}{\fancyhf{}\fancyhead[L]{Accepted for publication in IEEE Transactions on Image Processing on 15 Dec. 2021}}

\begin{document}
%
\title{End-to-End Rate-Distortion Optimized Learned \\ Hierarchical Bi-Directional Video Compression}
%
%
%

\author{M. Akın~Yılmaz,~\IEEEmembership{Member,~IEEE,}
     and~A.Murat~Tekalp,~\IEEEmembership{Fellow,~IEEE}

\thanks{Authors are with the Department of Electrical and Electronics Engineering, Koç University, 34450 Istanbul, Turkey.}
\thanks{This work was supported by TUBITAK 1001 project 217E033 and TUBITAK 2247-A Award No. 120C156. A.M.T. also acknowledges support from Turkish Academy of Sciences (TUBA).}
\thanks{This is a significantly expanded and rewritten version of a paper presented in ICIP 2020.}
\thanks{Manuscript submitted July 8, 2021, revised October 18, 2021.}}

\maketitle

\thispagestyle{head}

\begin{abstract}
Conventional video compression (VC) methods are based on motion compensated transform coding, and the steps of motion estimation, mode and quantization parameter selection, and entropy coding are optimized individually due to the combinatorial nature of the end-to-end optimization problem. Learned VC allows end-to-end rate-distortion (R-D) optimized training of nonlinear transform, motion and entropy model simultaneously. Most works on learned VC consider end-to-end optimization of a sequential video codec based on R-D loss averaged over pairs of successive frames. It is well-known in conventional VC that hierarchical, bi-directional coding outperforms sequential compression because of its ability to use both past and future reference frames. This paper proposes a learned hierarchical bi-directional video codec~(LHBDC) that combines the benefits of hierarchical motion-compensated prediction and~end-to-end optimization. Experimental results show that we achieve the~best R-D results that are reported for learned VC schemes to date in both PSNR and MS-SSIM. Compared to conventional video codecs, the R-D performance of our end-to-end optimized codec outperforms those of both x265 and SVT-HEVC encoders (``veryslow" preset) in PSNR and MS-SSIM~as~well~as HM 16.23 reference software in MS-SSIM. We present ablation studies showing performance gains due to proposed novel tools such as learned masking, flow-field subsampling, and temporal flow vector prediction. 
The~models and instructions to reproduce our results can be found in \url{https://github.com/makinyilmaz/LHBDC/}.
\end{abstract}

%
\IEEEpeerreviewmaketitle

\section{Introduction}
%
%
%
%

\IEEEPARstart{S}{tandards} based video compression has been based on the hybrid coding technology of block-based DCT compression of pixel/motion-compensated residual data since early 1980s. There has been a new video compression standard nearly every 10 years since the first broadcast quality video compression standard MPEG-2 in 1994. However, each new standard is improving the rate-distortion (R-D) performance over the former by introducing more encoding options at the cost of significantly higher complexity. Furthermore, selection of the optimal encoding configuration (in the R-D sense) is a combinatorial problem, which can only be addressed by heuristic solutions. Another limitation of traditional video compression methods is that they employ a linear transform, which may not be effective enough to reduce the entropy of latent variables to be compressed. There is also no easy method to optimize the entropy model for the arithmetic encoder. 

The study of learned image/video compression methods is motivated by two main effects: i) Exploiting the power of nonlinear transforms for decorrelation and reduction of the entropy of the latent variables, and ii) enable end-to-end optimization of the nonlinear transform, the motion-compensation (MC) model, and the entropy model simultaneously by gradient-based training. The latter converts the tedious and heuristic task of encoder optimization into a data-driven learning problem. As a result, the inference process of optimized learned image/video encoders is faster than traditional R-D optimization of image and video encoders.

In this work, we propose a novel learned hierarchical bi-directional video compression (LHBDC) framework that combines the~advantages of traditional hierarchical bi-directional MC with data-driven end-to-end R-D optimization. We introduce innovations, such as flow-field subsampling, flow-vector prediction, learned bi-directional MC masking, which we show, by means of ablation studies, that each contributes to the superior performance of our method. 
Our extensive experimental results demonstrate we achieve the best R-D results that are reported for learned video codecs to date in both PSNR and MS-SSIM. In addition, the R-D performance of our end-to-end optimized codec outperforms those of both x265 and SVT-HEVC encoders (``veryslow" preset) in PSNR and MS-SSIM~as~well~as HM 16.23 reference software in MS-SSIM.
Section~\ref{related} discusses related work and our contributions. The~proposed LHBDC framework is presented in Section~\ref{sec:method}. Experimental results are shown in Section~\ref{sec:experiment}. Finally, Section~\ref{sec:conclude} concludes the paper.


\section{Related Work and Contributions}
\label{related}
The state-of-the-art in learned image compression is based on the variational auto-encoder framework, introduced in the seminal work~\cite{balle_end}, for end-to-end R-D optimized image compression. Following this work, further improvements in the R-D performance have been obtained by means of more accurate but more complex entropy modeling~\cite{balle_scale,minnen_joint,gmm_cheng,minnen_channel}.

Existing end-to-end learned video codecs replace the functional blocks in a standards-based video codec with deep networks, such that different networks for key frame compression, motion estimation, motion compensation, motion vector compression, and residual compression are optimized together based on a single R-D loss. These end-to-end learned video codecs use fully convolutional and/or recurrent architectures, and they can be employed for sequential or bi-directional motion-compensated video coding. In the following, we review selected works that define the state of the art in learned video compression. DVC~\cite{dvc} is the first end-to-end deep video compression model that jointly learns all components of the video compression framework with promising results. For motion compensation, DVC employs flow-based back-warping, which inevitably produces artefacts in areas with fast motion or occlusion. This problem is alleviated by introducing a motion compensation post-processing network. Scale-space flow~\cite{scale_space} replaces the motion-compensation network in DVC by scale-space warping, where a scale channel is added to the flow-field tensor as a third dimension. The scale channel acts as an uncertainty parameter, which introduces blurring in regions that are prone to artefacts to yield a better inter-frame prediction and a residual with lower entropy.
DVC and scale-space flow methods compress the current frame with only one reference frame that limits their ability to fully exploit long term correlations in the video. In order to benefit from long-term correlations, Recurrent Learned Video Compression (RLVC)\cite{rlvc} proposes a Recurrent Auto-Encoder (RAE) and Recurrent Probability Model (RPM) for low-delay video compression with sequential motion compensation.

Hierarchical learned video compression (HLVC)~\cite{hlvc} is one of the first methods that treat end-to-end optimization of bi-directional inter-frame prediction concurrently with our method~\cite{our_icip_2020}. These two works were developed independently.

The main contributions of this paper are: \vspace{-1pt}
\begin{itemize} 
    \item We present a learned end-to-end optimized hierarchical bi-directional video coding framework that outperforms the SVT-HEVC encoder in PSNR and MS-SSIM and HM~16.23 reference codec in MS-SSIM for the first time.
    \item We train a single bi-directional motion-compensated compression network that works for all levels of the hierarchy to simplify decoder operation.
    \item We propose novel tools such as motion field subsampling, temporal bi-directional prediction of motion vectors, and learned bi-directional motion compensation mask, and show that each one contributes to the superior performance of our codec by ablation studies in Section~\ref{ablation}.
\end{itemize}


\section{LHBDC: End-to-End Optimized Learned Hierarchical Bi-Directional Video Compression}
\label{sec:method}
We formulate a learned hierarchical bi-directional video compression framework, where we can perform end-to-end R-D optimization. The proposed method is an extension of our initial work presented at the 2020 IEEE Int. Conf. on Image Processing (ICIP)~\cite{our_icip_2020}. The codec presented here has superior R-D performance compared to that reported in~\cite{our_icip_2020}. Section~\ref{overview} provides an overview of the~proposed video codec. Detailed explanation of each component and the end-to-end R-D optimization strategy are presented in the following. 

\subsection{System Overview}
\label{overview}

\begin{figure}[!ht]
	\centering
 \includegraphics[width=0.98\linewidth]{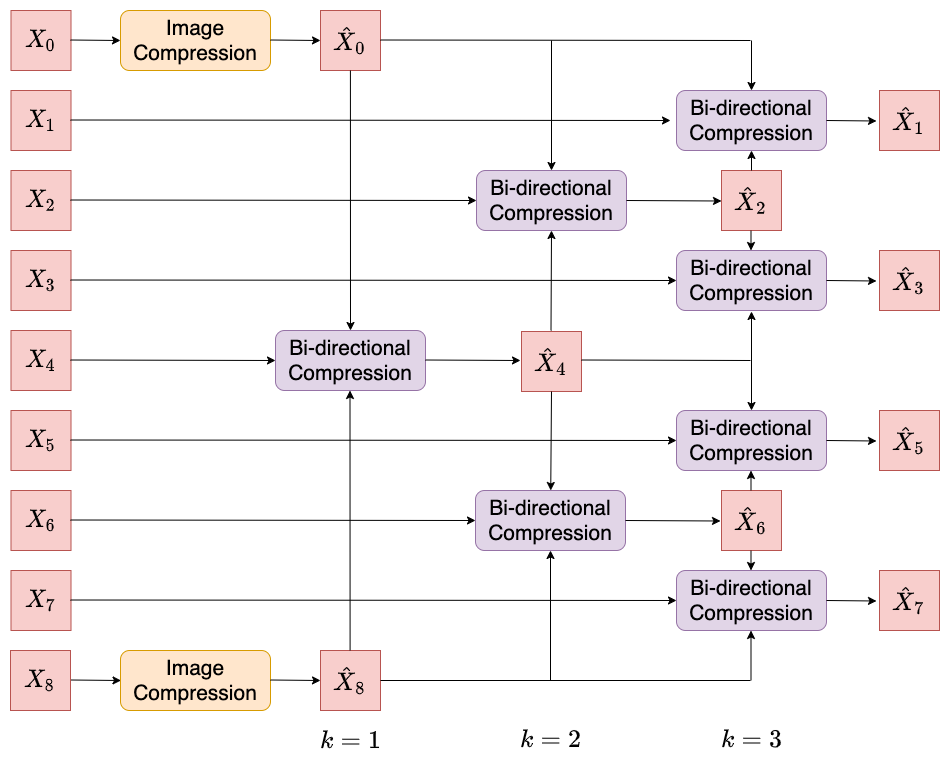}  \vspace{-14pt}
    \caption{Block diagram of $K=3$ level learned hierarchical bi-directional video codec (LHBDC). The~same colored blocks are identical networks that share learnable parameters. The pink squares denote video frames.}
    \label{fig:overall_diagram}  \vspace{-13pt}
\end{figure}

The architecture for the proposed end-to-end learned hierarchical bi-directional video compression (LHBDC) framework mimics the structure of the traditional hierarchical B-pictures video compression, replacing classical blocks for motion estimation, compensation, and compression and residual compression with trainable deep networks. Figure~\ref{fig:overall_diagram} illustrates the~block diagram for the overall LHBDC system with $K=3$ level hierarchy corresponding to group-of-pictures (GOP) with $8$~frames. The same procedure is applied for each GOP in a video sequence. There are two types of frames in the proposed system: keyframes and bi-directional predicted frames. 

The first frame of each GOP is taken as a keyframe, which is compressed using a simplification of the learned still-image compression method~\cite{gmm_cheng}. We simplified the codec in~\cite{gmm_cheng} by excluding the attention layers and using an entropy model with a single Gaussian for keyframe compression. We found these simplifications do not significantly affect the R-D performance but provide better computational efficiency.

\begin{figure}[!ht]
	\centering
    \includegraphics[width=1\linewidth]{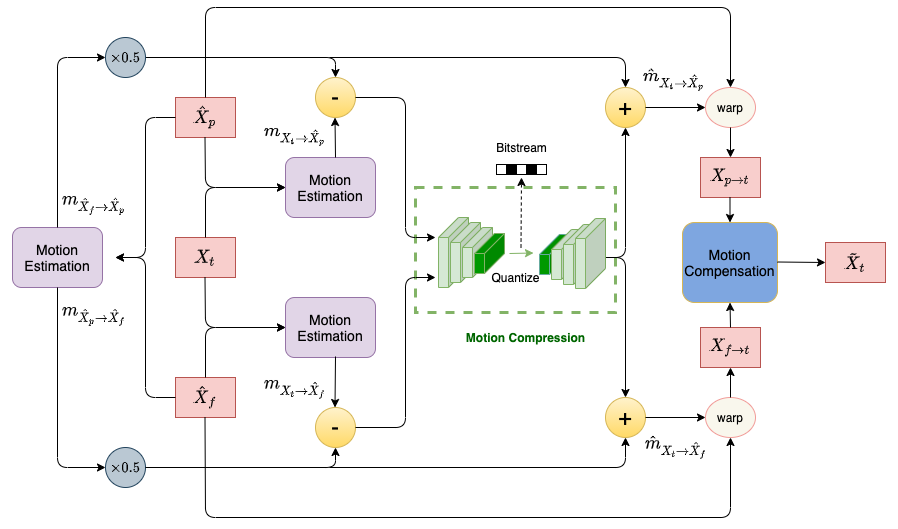} \vspace{-12pt} \\
	(a) \\
	\includegraphics[width=0.75\linewidth]{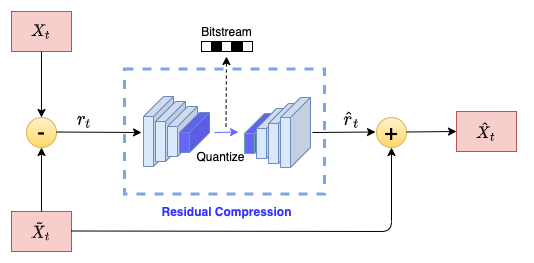} \vspace{-5pt} \\
	(b) \vspace{-5pt}
    \caption{Block diagram of the proposed learned bi-directional compression network (one of the~purple blocks in Figure~\ref{fig:overall_diagram}). (a)~Motion estimation, encoder-decoder and compensation networks, (b)~Residual encoder-decoder network.}
    \label{fig:bidirec_comp_diagram}
\end{figure}

All other frames are bi-directionally compressed, similar to the classical hierarchical B-pictures coding, using the proposed learned bi-directional compression network shown by purple blocks in Figure~\ref{fig:overall_diagram}. The keyframes are used as forward and backward references for motion compensation of the~middle frame of the GOP. For all other frames the~nearest available decoded past and future frames are used as backward and forward references. For $K=3$ level hierarchy, we run~7~instances of the proposed network, which take different frame pairs as input but share the same learnable parameters (see Figure~\ref{fig:overall_diagram}).

The proposed bi-directional compression network (the inside of purple box in Fig.~\ref{fig:overall_diagram}) is composed of 
motion estimation, motion compensation, motion field compression, and residual compression subnetworks, which are shown in Figure~\ref{fig:bidirec_comp_diagram}~(a) and (b), respectively. 
Motion field compression module features motion subsampling and temporal motion prediction to increase the efficiency of motion compression. Motion field subsampling is analogous (resolution-wise) to sending one motion vector per 4x4 block of pixels. Temporal motion prediction relies on constant velocity motion assumption to reduce the dynamic range of vectors to be compressed. There are some inevitable motion compensation artifacts at motion boundaries due to occlusion effects even if we do not employ motion subsampling. We propose the learned bi-directional motion compensation network, which is optimized to learn the best continuous coefficients to merge the forward and backward warped frames, to mitigate such motion artifacts. These components are explained in detail in Section~\ref{Motion_compression} to Section~\ref{Residual_compression}. Their effectiveness is demonstrated by ablation studies in Section~\ref{ablation}.

\subsection{Bi-directional Motion Estimation and Compression}
\label{Motion_compression}
Motion (flow) field refers to the set of all forward or backward motion vectors for a video frame. We first discuss forward and backward motion field estimation and then introduce our motion-field compression subnet. 

\subsubsection{Hierarchical Dense Flow (Motion) Field Estimation}
We~adopted the spatial pyramid network (SpyNet)~\cite{spynet} as our flow estimation subnet, since it handles large motion vectors well. 
In order to generate a prediction for the target (current) frame $X_t$ by bi-directional motion compensation, we estimate the flow field $m_{X_t \rightarrow \hat{X}_p}$ from the~target frame $X_t$ to the decoded past reference frame $\hat{X}_p$  as well as the flow field $m_{X_t \rightarrow \hat{X}_f}$ from $X_t$ to the decoded future reference frame $\hat{X}_f$. 
For example, for level $k=1$ we estimate the motion fields $m_{4\rightarrow \hat{0}}$ and $m_{4\rightarrow \hat{8}}$, which are then encoded by the~motion compression subnet and transmitted to the decoder.

In addition, level $k=1$ bi-directional compression network also estimates flow fields $m_{\hat{0}\rightarrow \hat{8}}$ and $m_{\hat{8}\rightarrow \hat{0}}$ given the~compressed and decoded key frames $\hat{X}_{0}$ and~$\hat{X}_{8}$. These flow fields are used in the temporal motion vector prediction step discussed in the next subsection. They are not compressed and transmitted since the decoder can also estimate them. Note that for other levels of the hierarchy (i.e., $k=2,3$), the flow field between the past and future reference frames in one direction is available from the previous hierarchy level.

\subsubsection{Motion Field Compression by Subsampling and Temporal Prediction}
We employ spatial subsampling of backward and forward motion fields and temporal prediction of motion vectors to reduce the bitrate for encoded motion vectors. 

The estimated backward and forward motion fields can be spatially subsampled using separable cubic interpolation filtering. The subsampled motion fields go through the motion encoder and decoder. The output of motion decoder is then upsampled and used to generate the backward and forward warped prediction frames in the bi-directional motion compensation step discussed in the next subsection.

In order to further reduce the entropy of motion fields to be encoded, we perform temporal prediction of subsampled motion vectors and encode the prediction errors. 
Motion vectors at hierarchy level $k$ are predicted by half of motion vectors between the past and future reference frames for level $k$. As stated above, for level  1, the bi-directional  motion vectors between the past and future reference frames 0 and 8 are estimated separately, and the difference vectors $m_{4\rightarrow \hat{0}} - 0.5 \times m_{\hat{8}\rightarrow \hat{0}}$ and $m_{4\rightarrow \hat{8}} - 0.5 \times m_{\hat{0}\rightarrow \hat{8}}$ are encoded.

\begin{figure}[b!]
	\centering
 \includegraphics[width=1\linewidth]{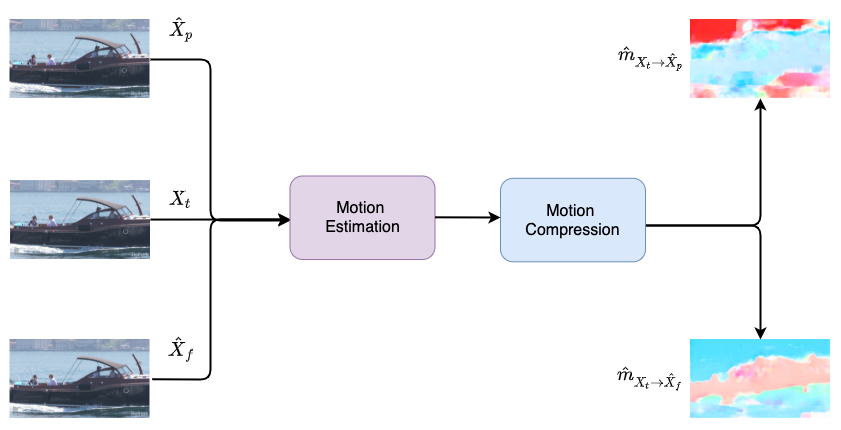} \vspace{-20pt}
    \caption{Example estimated bi-directional flow field (magnitude) after~motion compression-decompression with optional motion down- and up-sampling.}
    \label{fig:decoded_bi_directional_flows_yatch5}
\end{figure}

We use a modified version of the keyframe compression network to compress the bi-directional flow difference tensor. The~input to motion compression subnet is a 4-channel tensor, with the shape $H \times W \times 4$, comprised of the backward and forward motion vector differences ($2$ channels for backward and $2$ channels for forward) stacked together. For the motion compression network, the number of convolution filters has been set to $N = 128$. 
Finally, the decoded bi-directional flow vector differences are added to the predicted motion vectors to get the reconstructed bi-directional motion fields.
A~visualization of the decoded flow vectors after bi-directional motion estimation and compression is shown in Figure~\ref{fig:decoded_bi_directional_flows_yatch5}.

\subsection{Bi-directional Motion Compensation}
\label{Motion_compensation}

\begin{figure}[t!]
	\centering
 \includegraphics[width=1\linewidth]{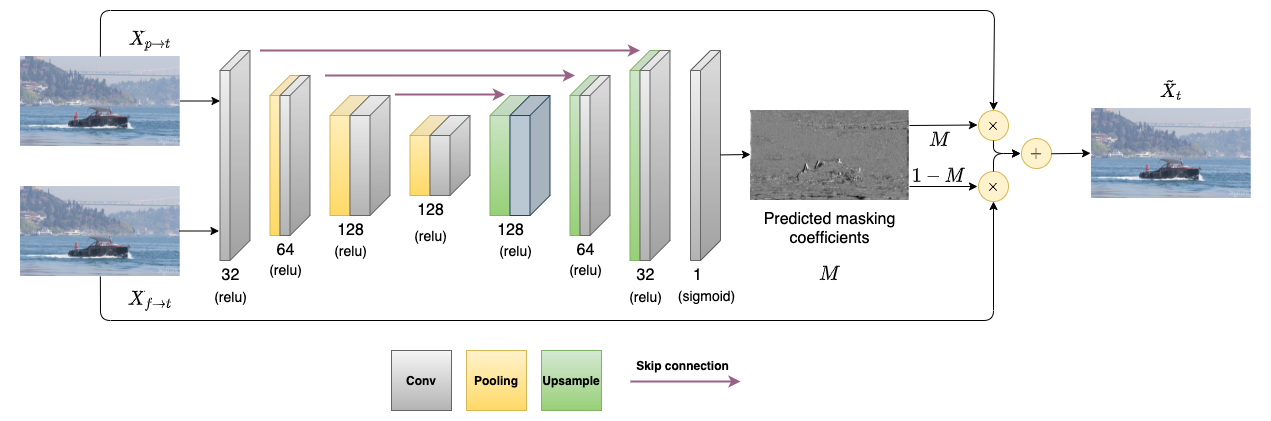} \vspace{-16pt}
    \caption{Block diagram of the U-shaped motion compensation network. The~output of the network is a mask image $M_t$ containing pixel-wise weights to fuse the~warped forward and backward reference frames.}
    \label{fig:motion_compensation_net}
\end{figure}

Bi-directional motion compensation is achieved by first warping the past and future reference frames $\hat{X}_p$ and $\hat{X}_f$ towards the current frame by bilinear mapping using the decoded backward and forward flow vectors $\hat{m}_{X_t \rightarrow \hat{X}_p}$ and $\hat{m}_{X_t \rightarrow \hat{X}_f}$ (shown in Figure~\ref{fig:decoded_bi_directional_flows_yatch5}), to form the forward and backward warped frames $\hat{X}_{p \rightarrow t}$ and $\hat{X}_{f \rightarrow t}$, respectively. 

The forward and backward warped frames $\hat{X}_{p \rightarrow t}$ and $\hat{X}_{f \rightarrow t}$ are input to a \textit{motion compensation mask estimation subnet}, shown in Figure~\ref{fig:motion_compensation_net}, which learns the pixel-wise weights (denoted by the mask image $M_t$) to fuse them in order to form a single motion-compensated frame with improved quality. The~network has a U-shaped architecture similar to that in~\cite{voxel} to estimate the pixel-wise mask $M_t$ to fuse $\hat{X}_{p \rightarrow t}$ and $\hat{X}_{f \rightarrow t}$. Sigmoid non-linearity is applied at the output layer of the~motion compensation mask network to ensure that the~mask values are between 0 and 1. 

Finally, the  fused motion-compensated prediction $\tilde{X}_{t}$ for the~current frame~$t$ is computed using the mask and forward and backward warped reference frames as
\begin{equation}
\tilde{X}_{t} = M_t \cdot X_{p\rightarrow t} + (1-M_t) \cdot X_{f\rightarrow t}
\end{equation}
where $M_t$ contains pixel-wise masking coefficients for frame~$t$. 


\subsection{Residual Compression}
\label{Residual_compression}

The residual between the motion compensated frame $\tilde{X}_{t}$ and the original frame is compressed by using a network with the same architecture as the keyframe compression network. Similar to the motion compression subnet, the number of filters is set to $N = 128$.  Decoded residual $\hat{r}_t$ is then added to the motion compensated frame at the decoder. The block diagram of the residual compression network and reconstruction of current frame $\hat{X}_t$ are shown in Figure~\ref{fig:bidirec_comp_diagram}. 


\subsection{End-to-End Rate-Distortion Optimization}
\label{e2e}
The proposed compression framework is optimized to minimize the number of bits to represent a GOP while maximizing the quality of decoded frames. To that effect, we define a R-D loss as  
\begin{equation}
L = \lambda D + R_{image} +  R_{flow} + R_{residual}
\end{equation}
where $D$ denotes the distortion between the original and reconstructed frames accumulated over a GOP. The mean square error (MSE) is used for the distortion measure. $R_{image}$, $R_{flow}$ and $R_{residual}$ stand for the rates in bits/pixel for the compressed keyframe, bi-directional flow field, and residual latents accumulated over a GOP. Since the~network architectures for keyframe, flow field, and residual compression are similar to that presented in~\cite{gmm_cheng}, which utilizes a hyper-prior network for entropy modeling, the~rates are estimated by the~entropy of the respective latents and hyper-prior latents.

The Lagrange multiplier $\lambda$ in the loss function determines a R-D trade-off point. Since $\lambda$ multiplies the distortion term, an increase in $\lambda$ leads to reducing distortion in exchange for an increase in the bit-rate. 
In order to generate a R-D curve, the proposed framework is optimized for a set of 4-6 $\lambda$ values, which are empirically selected in order to cover a bit-rate range of interest. Note that each different~$\lambda$ value leads to a different end-to-end optimized network model. Hence, learned video encoding at different R-D points is implemented using different learned models (with fixed quantization~\cite{balle_nonlin}) in the same spirit as encoding-decoding videos with different quantization parameter $QP$ values in conventional video encoding.

\subsection{Decoder Operation}
\label{decoder}
We discussed the decoder indirectly up to now, since the~encoder includes a copy of the decoder in video compression to prevent drift. Here, we summarize the operation of a stand alone decoder, since the complexity and efficiency of the decoder are important to operate the decoder with an acceptable image resolution and frame rate in real-time.

\begin{figure}[t!]
	\centering
 \includegraphics[width=1\linewidth]{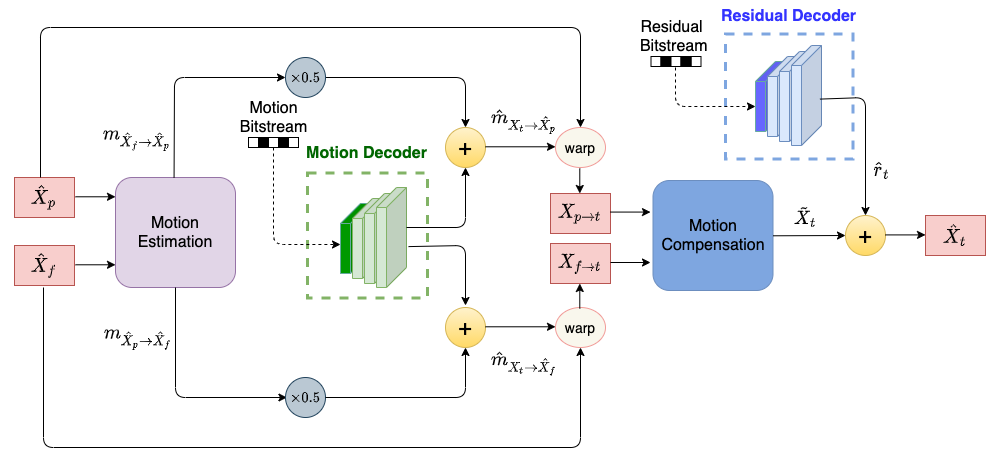} \vspace{-18pt}
    \caption{Block diagram of the decoder. Motion estimation between decoded past and future reference frames is used for temporal motion vector prediction. }
    \label{fig:proposed_decoder}
\end{figure}

The decoder takes the motion field and motion-compensated residual latent bitstreams for the current frame $X_t$ and produces a reconstructed frame $\hat{X}_t$ given the decoded past reference $\hat{X}_p$ and decoded future reference frame $\hat{X}_f$ in the frame buffer. The decoder network is shown in Figure~\ref{fig:proposed_decoder}. 

The decoder in the proposed codec must perform motion estimation between the past and future reference frames as part of temporal motion vector prediction. We note that bi-directional motion estimation is performed for layer 1 in the hierarchy, while only unidirectional motion estimation from the past to the future reference frame needs to be done for all decoders at all other hierarchy levels. This is because motion vectors from the future to the past reference for level $k$ are already computed in the decoder for level $k-1$.

\section{Experimental Results}
\label{sec:experiment}
We discuss the datasets, training details and evaluation methodology in Section~\ref{train}. The R-D performance of the proposed end-to-end optimized codec is compared with the state of the art learned and conventional codecs in Section~\ref{perf}. Finally, in Section~\ref{ablation} we perform an ablation study to analyze the contribution of each subnetwork and some parameter choices to the performance of the overall system.

\subsection{Datasets, Training Details, and Evaluation Methodology}
\label{train}
Deep learning framework PyTorch~\cite{pytorch} is used for implementing the proposed learned hierarchical bi-directional video compression method. We set the default GOP size to $N=8$ since it needs to be a power of~2 in~the hierarchical framework.

For keyframe compression, we use the mean-scale hyper-prior image compression model~in the CompressAI library~\cite{compressai}  (see \url{https://interdigitalinc.github.io/CompressAI/}).  We employ the~pre-trained models ''mbt2018$\_$mean" for $5$~$\lambda$ values $(0.0130, 0.0250, 0.0483, 0.0932, 0.1800)$ referring to quality levels 4, 5, 6, 7, 8 to compress keyframes at different rates. 
We use the nearest integer quantization as in~\cite{balle_nonlin}.

For forward and backward motion estimation, we use the pre-trained SPyNet model (see \url{https://github.com/anuragranj/spynet}) because of its real-time performance and efficiency. 

The proposed bi-directional compression model is trained using the Vimeo septuplet dataset~\cite{vimeo}. The septuplet dataset contains 91,701 7-frame sequences with fixed resolution $448 \times 256$. 
We trained a single model for bidirectional compression at hierarchy levels 1, 2 and 3. To this effect, we form sub-sequences of three pictures each, consisting of frames (1-4-7), (1-3-5 or 2-4-6 or 3-5-7), and (1-2-3 or 2-3-4 or 3-4-5 or 4-5-6 or 5-6-7), emulating level 1, level~2 and level 3 of the hierarchy, respectively, from the~septuplet dataset. They are randomly cropped to size $256 \times 256$ to implement data augmentation. We set the~mini-batch size to 4 and randomly select 4 sub-sequences from the set of all sub-sequences to train the model for compression of the middle picture in each iteration. Note that levels 1, 2, and 3 generate subsequences with different motion ranges; hence, the network is well-trained to cover a variety of motion ranges.

The motion estimation, motion compression, motion compensation mask estimation, and residual compression networks are trained jointly with a single end-to-end R-D loss. The~motion compensation mask estimation network, as well as motion residual and frame residual compression networks are initialized at random, while the motion estimation subnet is initialized with the~pre-trained SpyNet model weights and the parameters of the key-frame compression network are frozen and not updated. Since $B$ frames contribute less to the overall bit-rate compared to keyframes, we found that using $\lambda$ values targeting lower bitrates for the bi-directional compression module compared to $\lambda$ values used for keyframes yield better overall R-D performance. Hence, we trained the bi-directional compression module using $\lambda$ values $(0.0035, 0.0067, 0.0130, 0.0250, 0.0483)$, which are tuned to lower bitrates. The model is trained for $1$ million iterations using ADAM~\cite{adam} optimizer. The initial learning rate is set to $10^{-4}$. The learning rate is reduced by a factor of $2$ when the~loss is not reduced for $25K$ iterations.

The evaluation of the proposed method is conducted on the UVG dataset~\cite{uvg}, which is commonly used in the literature for comparison of different learned video codecs. The dataset contains $50/120$ fps, $1080$p and $4K$ videos. As in the literature, the comparisons are conducted on $7$ different $1080$p videos: \textit{Beauty, Bosphorus, Honeybee, Jockey, ReadySetGo, ShakeNDry} and \textit{YachtRide}, which contain different motion characteristics. 
There are multiple ways to compute the average distortion for a set of videos~\cite{psnr_computation}. We report the average of PSNR/MS-SSIM of individual frames over the whole dataset in order to be able to compare our results with those of other learned codecs, which report the same.
We~also report the~average R-D performance for each video separately.

\subsection{Compression Performance Comparison}
\label{perf}
We compare the R-D performance of our codec, in terms of both PSNR and MS-SSIM distortion measures, to those of other leading learned sequential video compression methods, such as DVC~\cite{dvc}, Scale-space flow~\cite{scale_space}) and RLVC~\cite{rlvc}, and bi-directional video compression method HLVC~\cite{hlvc}, as well as some open source traditional H.265/HEVC video codecs~\cite{h265} (both low-latency and hierarchical modes). The R-D curves of the aforementioned learned codecs are taken as is from their respective github pages.

We selected three opensource H.265/HEVC encoders as anchors: the low-latency x265 (veryslow preset), the hierarchical B-frame compression mode of SVT-HEVC codec with the medium (default) and veryslow presets, and the HM~16.23 reference software. Experiments are conducted~following the~configurations described in Appendices~\ref{appendixA},~\ref{appendixB} and~\ref{appendixC}. 

We~provide average R-D curves and average Bjøntegaard Delta Bit-Rate (BD-BR) performance comparison of different codecs over 7 videos in the UVG testset in  Section~\ref{rdcurve}. The~BD-BR performance of our codec compared to SVT-HEVC codec (hierarchical mode) with medium and veryslow presets for individual videos is presented in Section~\ref{bdrate}.

\subsubsection{Rate-Distortion Evaluation in terms of PSNR}
\label{rdcurve}

\begin{figure}[!b]
	\centering
 \includegraphics[width=1\linewidth]{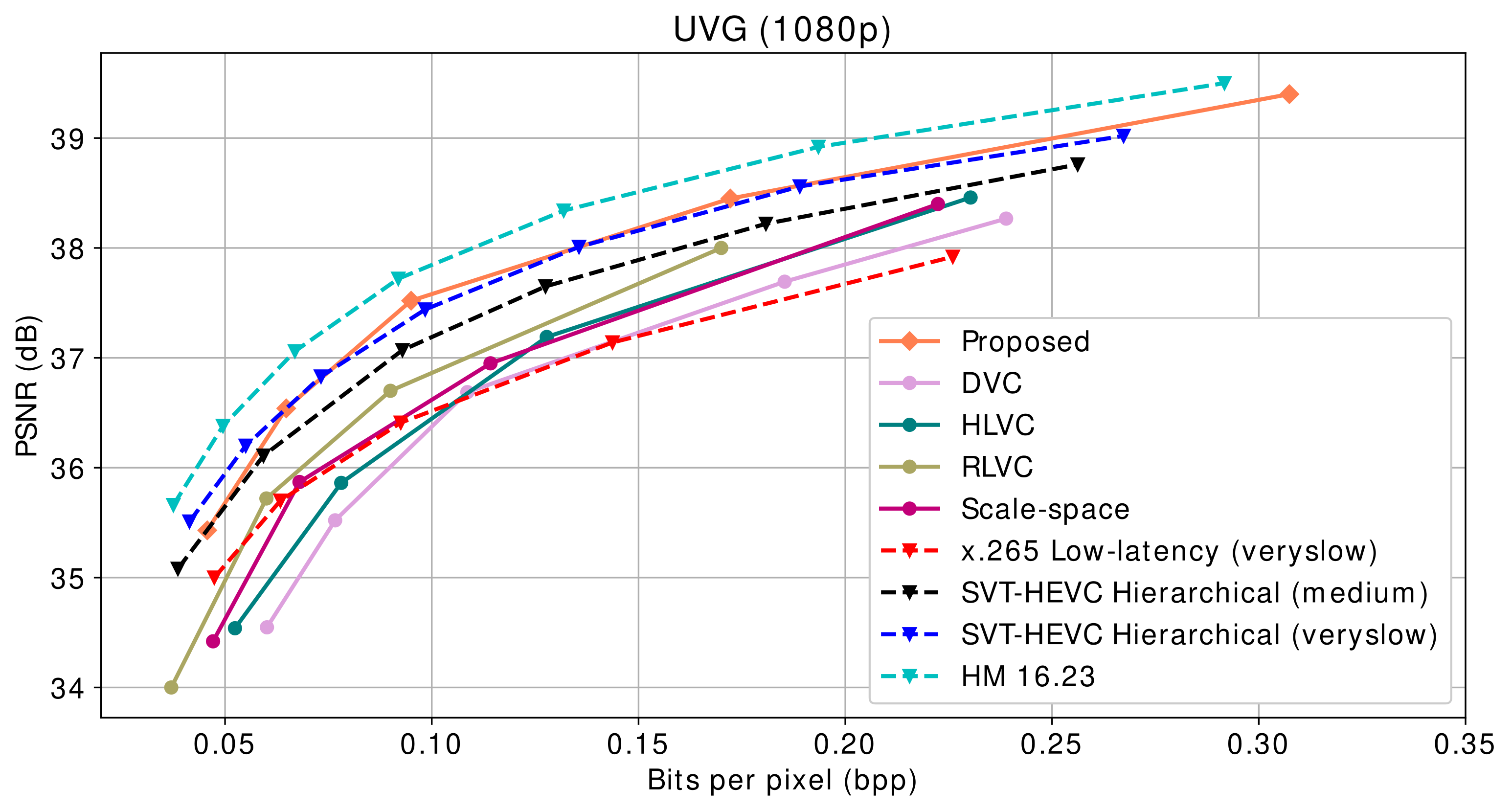} \vspace{-20pt}
    \caption{R-D performance comparison on the UVG test set using PSNR.}
    \label{fig:rd_curve_psnr}
\end{figure}

\begin{figure}[!t]
	\centering
 \includegraphics[width=1\linewidth]{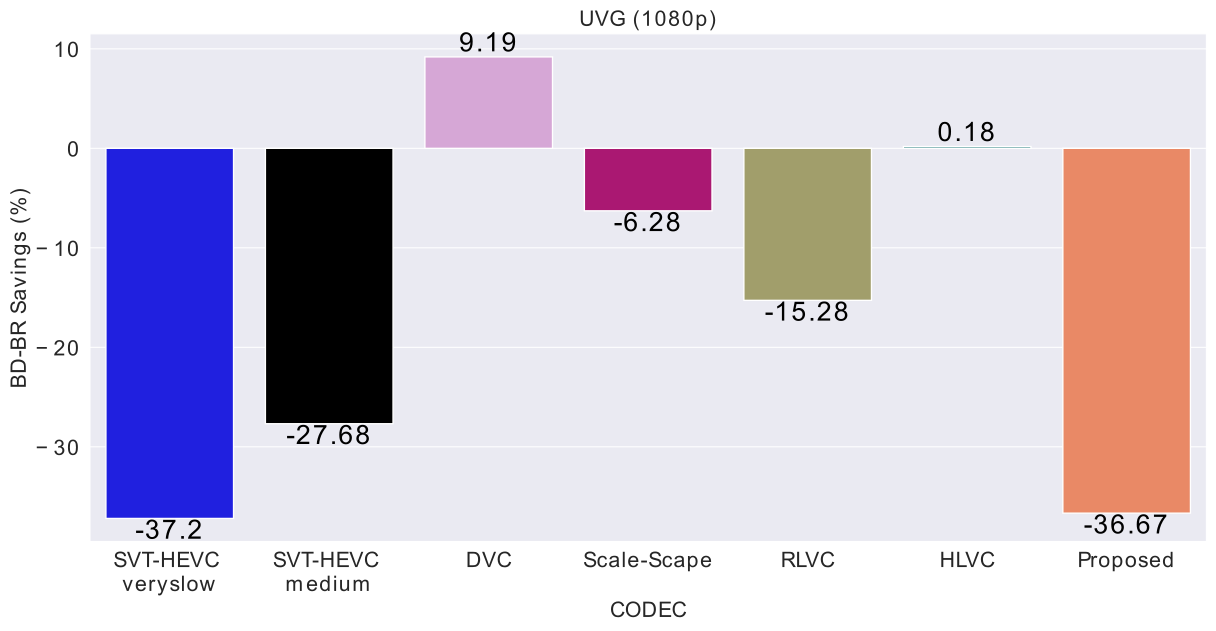} \vspace{-20pt}
    \caption{BD-BR reduction of different codecs including the proposed codec over the anchor x265 encoder (veryslow preset) on the UVG test set~\cite{uvg}.}
    \label{fig:proposed_bdbr_fig}
\end{figure}

The average R-D curves for different codecs, where the rate and and distortion are measured in bits per RGB pixel (bpp) and peak signal-to-noise ratio (PSNR) in decibels, respectively, over the~UVG dataset are shown in Figure~\ref{fig:rd_curve_psnr}. The R-D curves are obtained by cubic interpolation of R-D points computed for 5-6 different $\lambda$ values. The four anchors, low-latency x265 veryslow preset, hierarchical SVT-HEVC veryslow and medium presets, and HM 16.23 are shown by red, blue, black and cyan dotted curves in Figure~\ref{fig:rd_curve_psnr}, respectively.

Examination of Figure~\ref{fig:rd_curve_psnr} shows that we achieve significant performance improvement over the state-of-the-art learned video codecs as well as x265 (veryslow preset) and hierarchical B-frame mode of SVT-HEVC (medium preset). We are also on par with the veryslow preset of SVT-HEVC. At present, we cannot exceed the PSNR performance of HM16.23 reference software (similar to all other published learned codecs). 
We believe we are the first to report a learned video codec that outperforms {\it the~medium preset} and is on par with the veryslow preset of SVT-HEVC hierarchical coding.

In order to compare the curves in Figure~\ref{fig:rd_curve_psnr} quantitatively, the BD-BR measure~\cite{bjontegaard} with respect to the veryslow preset of x265 codec (the lowest quality anchor) is evaluated. This metric computes an average bit-rate reduction as the percent difference in rate with respect to the anchor at the same quality level over a range of quality values. Figure~\ref{fig:proposed_bdbr_fig} shows the average bit-rate reduction of hierarchical SVT-HEVC, DVC, HLVC, RLVC, Scale-space Flow and the proposed codecs over the low-latency x265 anchor on the UVG dataset. Negative numbers indicate the BD-BR reduction of each codec compared to the anchor. It can be calculated that our proposed codec yields near $13 \%$ BD-BR reduction compared to hierarchical mode of SVT-HEVC medium preset and is on par for veryslow preset. As can be seen from the figure, our codec also has superior BD-BR performance over all other learned video codecs, including the hierarchical bi-directional HLVC codec. 

\subsubsection{Rate-Distortion Evaluation in terms of MS-SSIM}
\label{rdmssim}
\begin{figure}[!t]
	\centering
 \includegraphics[width=1\linewidth]{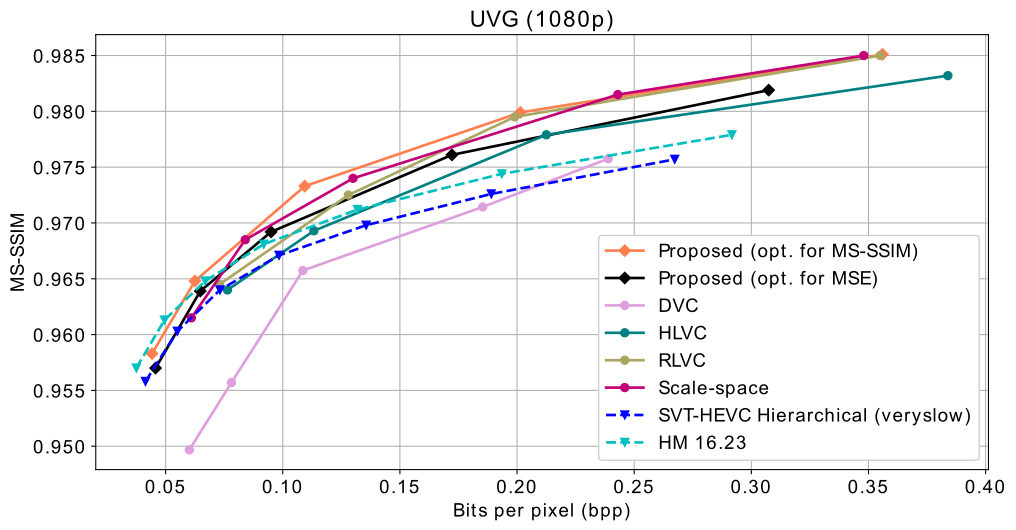} \vspace{-20pt}
    \caption{R-D performance comparison on the UVG test set using MS-SSIM.}
    \label{fig:rd_curve_ssim}
\end{figure}

Comparison of the R-D performance in terms of the MS-SSIM distortion measure, depicted in Fig.~\ref{fig:rd_curve_ssim}, shows that we (both our PSNR and MS-SSIM optimized learned codecs) handily outperform all standards based codecs including the HM~16.23 reference software, while the performance of learned scale-space flow and RLVC codecs are close to our performance especially at high bitrates.

\subsubsection{BD Bit-Rate Comparison for Individual Videos}
\label{bdrate}

\begin{figure}[!b]
	\centering
 \includegraphics[width=1\linewidth]{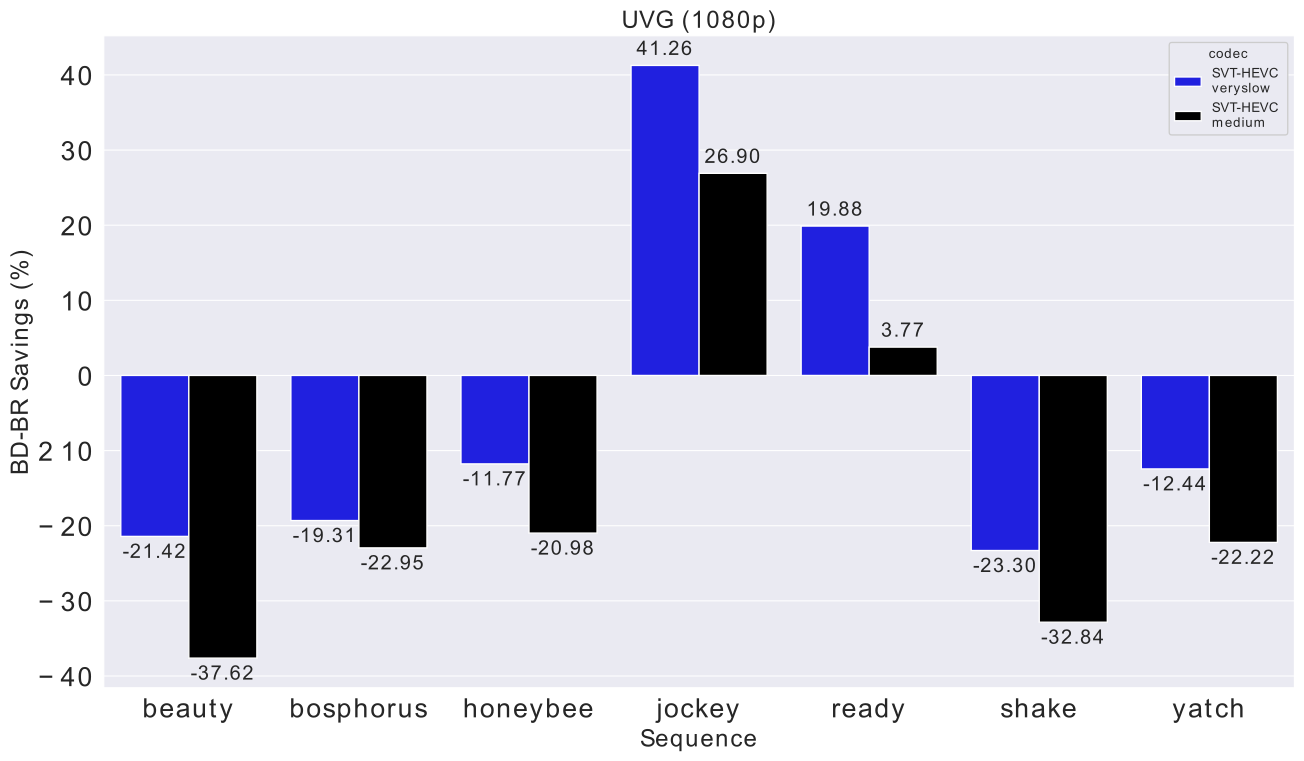} \vspace{-20pt}
    \caption{BD-BR reduction of our learned codec over SVT-HEVC hierarchical coding with medium (black) and veryslow (blue) presets on the UVG testset.}
    \label{fig:proposed_bdbr_sequence_fig}
\end{figure}

We analyze the BD-BR performance of our proposed codec for each video sequence individually in Figure~\ref{fig:proposed_bdbr_sequence_fig}. Since the~BD-BR values for each video sequence individually are not reported in other learned video codecs and the data is not available in their github pages, only our per sequence BD-BR results computed versus the hierarchical mode of SVT-HEVC codec with GOP size $N=2^3=8$ are presented. 

As can be seen from the figure, our proposed codec yields significant BD-BR reduction vs. SVT-HEVC (medium and veryslow presets) for all video sequences except for \textit{jockey} and \textit{ready}. These two sequences contain fast and complex motion, which results in motion-compensation artifacts that appear to be expensive to encode. Note that the proposed codec achieves more than $37\%$ and $21\%$ BD-BR reductions compared to medium and veryslow presets of hierarchical SVT-HEVC coding for the beauty sequence, which is a close-up with moderate texture.

\subsection{Ablation Study and Model Analysis}
\label{ablation}
The default configuration of our codec has three-level hierarchy (GOP size 8) with a intra-coded keyframe at every 8 frames, motion/flow field subsampling by a factor of 4 in each direction, temporal flow prediction, and no backward adaptive context model. The last choice is motivated by having a reasonable decoding time per frame, although we show below that having a backward adaptive context model indeed increases the R-D performance. 

In~this section, we investigate how much each proposed new component and how the choice of some parameters affect the~R-D performance of the proposed codec. 

\subsubsection{The Effect of Motion Field Subsampling}
\begin{figure}[!t]
	\centering
 \includegraphics[width=1\linewidth]{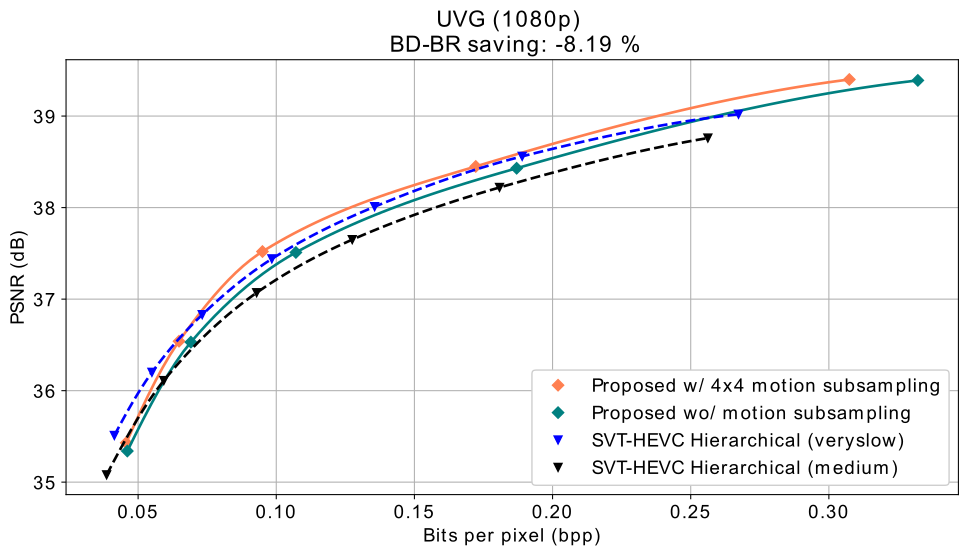}  \vspace{-19pt}
    \caption{Effect of proposed motion/flow field subsampling on the R-D curve.}
    \label{fig:proposed_rd_curve_subsample}
\end{figure}

We start by investigating how motion field subsampling affects the~R-D performance of the proposed method. The average R-D curves with and without motion subsampling in Figure~\ref{fig:proposed_rd_curve_subsample} show that subsampling the motion field by a factor of 4 in both directions provides $8.19\%$ BD-BR reduction. Note that motion/flow field is subsampled for compression only and we interpolate the subsampled motion fields back to full size before they are used for motion compensation. 

\begin{figure}[!t]
	\centering
 \includegraphics[width=0.85\linewidth]{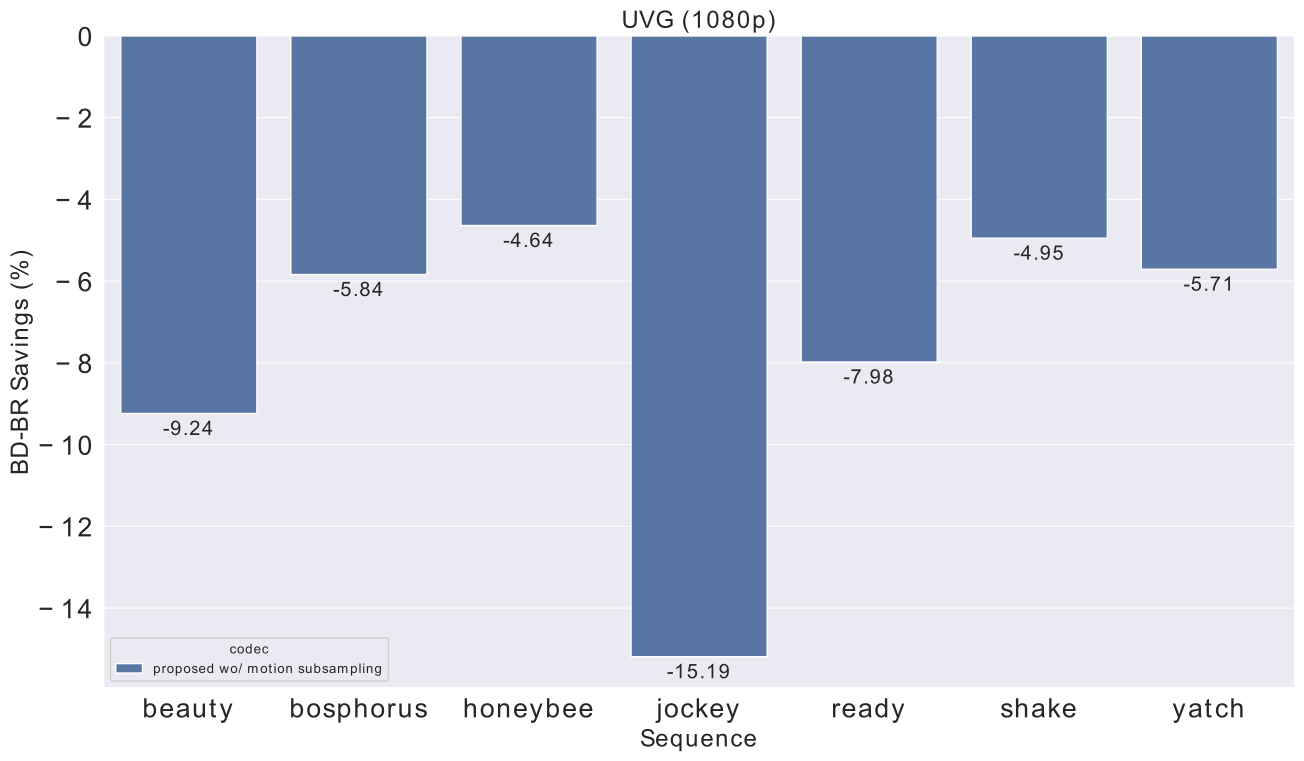} \vspace{-7pt}
    \caption{BD-BR reduction of our codec with vs. without motion subsampling.}
    \label{fig:proposed_bdbr_sequence_motion_sub}
\end{figure}

This decimation followed by interpolation results in smoothing of flow fields and is similar in essence to 4 $\times$ 4 block motion compensation in traditional codecs. Inspection of Figure~\ref{fig:proposed_bdbr_sequence_motion_sub} shows that motion field subsampling helps in the case of both videos with slow, predictable motion such as \textit{beauty} and videos with multiple motions and occlusions, such as \textit{jockey}.

\subsubsection{The Effect of Temporal Flow Prediction}
\begin{figure}[!t]
	\centering
 \includegraphics[width=1\linewidth]{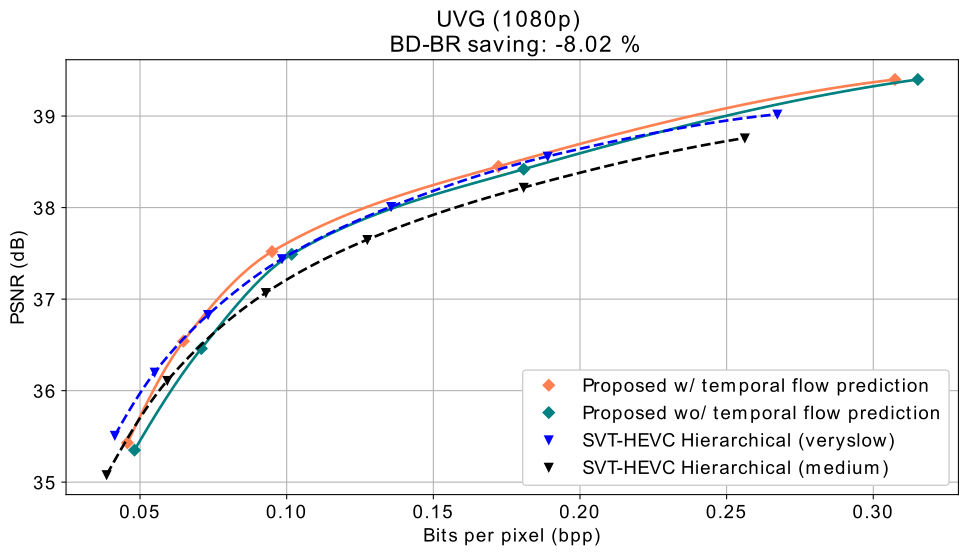}\vspace{-8pt}
    \caption{Effect of proposed temporal flow prediction on the RD-curve}
    \label{fig:proposed_rd_curve_flow_pred}
\end{figure}

We next investigate the effect of temporal flow prediction on the R-D performance of our codec. The R-D curves with temporal motion prediction on and off are depicted in Figure~\ref{fig:proposed_rd_curve_flow_pred}. The~R-D curves for the hierarchical SVT-HEVC codec with GOP size 8 medium and veryslow presets are also included for comparison. The~average R-D curves with and without temporal flow prediction in Figure~\ref{fig:proposed_rd_curve_flow_pred} show that motion field prediction using the previously decoded future and past reference frames provides $8.02\%$ BD-BR reduction.

\begin{figure}[!t]
	\centering
 \includegraphics[width=0.87\linewidth]{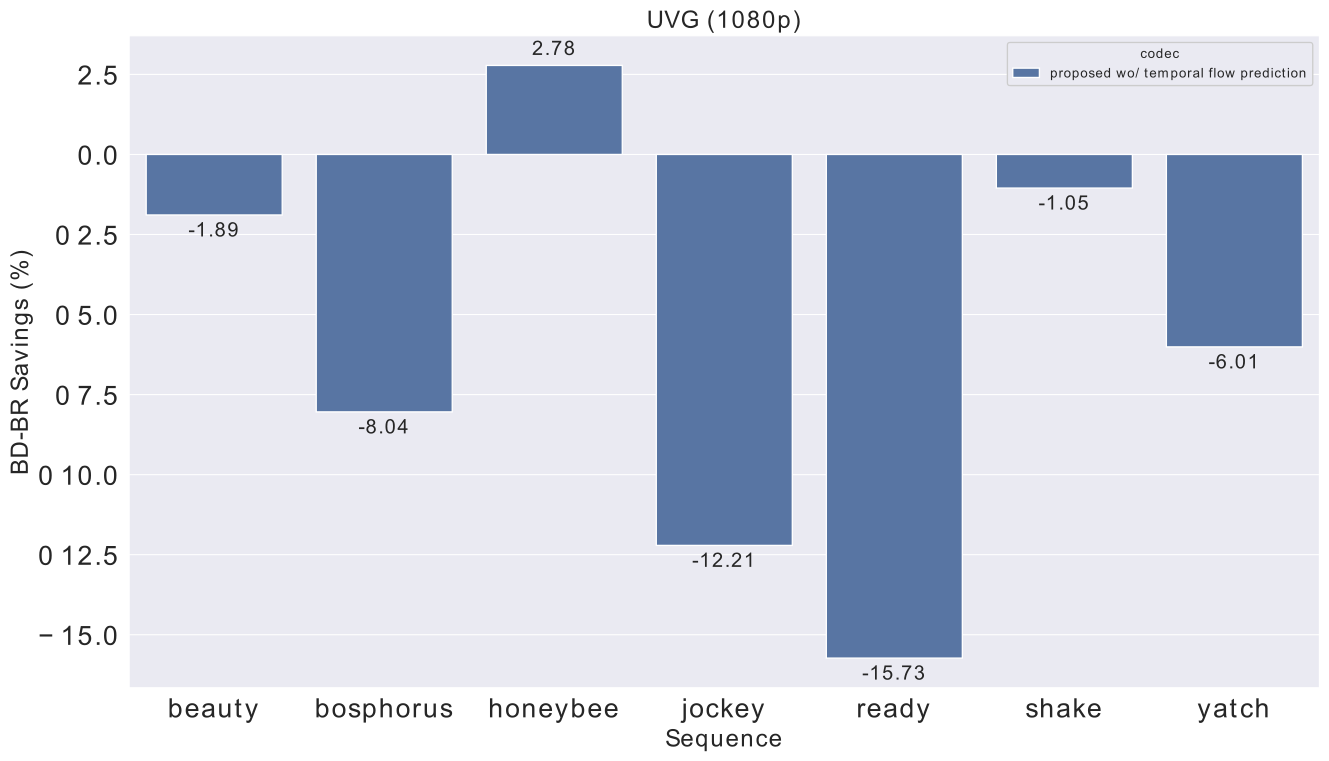} \vspace{-8pt}
    \caption{BD-BR reduction of our codec with vs. without temporal flow prediction.}
    \label{fig:proposed_bdbr_sequence_tfp}
\end{figure}

Figure~\ref{fig:proposed_bdbr_sequence_tfp} depicts BD-BR values using our codec with temporal flow prediction vs. no temporal prediction for individual videos. Temporal flow prediction has brought significant gains in videos with complex large motions, such as \textit{jockey, ready}, where taking the difference of estimated and predicted flow results in a decrease in the dynamic range of flow vectors to be compressed. In addition, in videos with global camera motion but no complex local motion, such as \textit{bosphorus, yatch}, where motion prediction is easier, we can see BD-BR gains due to temporal flow prediction. We see that in sequences with random, unpredictable motion, such as \textit{honeybee, shake}, temporal flow prediction may not provide gains.

\subsubsection{The Effect of Learned Motion Compensation Mask}
It is well-known that bi-directional block motion compensation (MC) with a simple averaging fusion yields better predicted frames compared to uni-directional MC in traditional video codecs. In this subsection, we demonstrate that bi-directional warping prediction, computed by fusing forward and backward flow-based warped frames using a learned pixel-wise mask, proposed in Section~\ref{Motion_compensation} as an alternative to simple averaging, yields better motion compensated predicted frames that are closer to the ground truth frames in learned video codecs. 

\begin{figure}[!ht]
	\centering
 \includegraphics[width=1\linewidth]{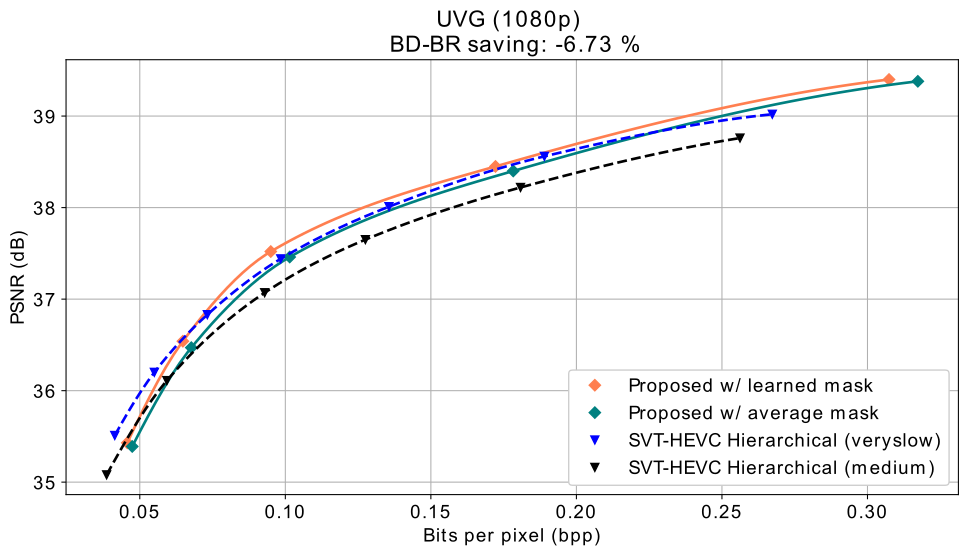} \vspace{-20pt}
    \caption{Effect of learned motion compensation on the R-D curve.}
    \label{fig:proposed_rd_curve_mc}
\end{figure}

The R-D curves to compare the performance of the proposed learned pixel-wise fusion mask vs. simple averaging are depicted in Figure~\ref{fig:proposed_rd_curve_mc}. It can be seen that the R-D curve with the learned fusion mask is above the one that is obtained by simple averaging of forward and backward warped frames over the entire bit-rate range and provides $6.73 \%$ BD-BR reduction, despite the fact that even the codec that employs simple averaging exceeds the performance of SVT-HEVC encoding with both medium and veryslow presets.

\begin{figure}[!ht]
	\centering
 \includegraphics[width=1\linewidth]{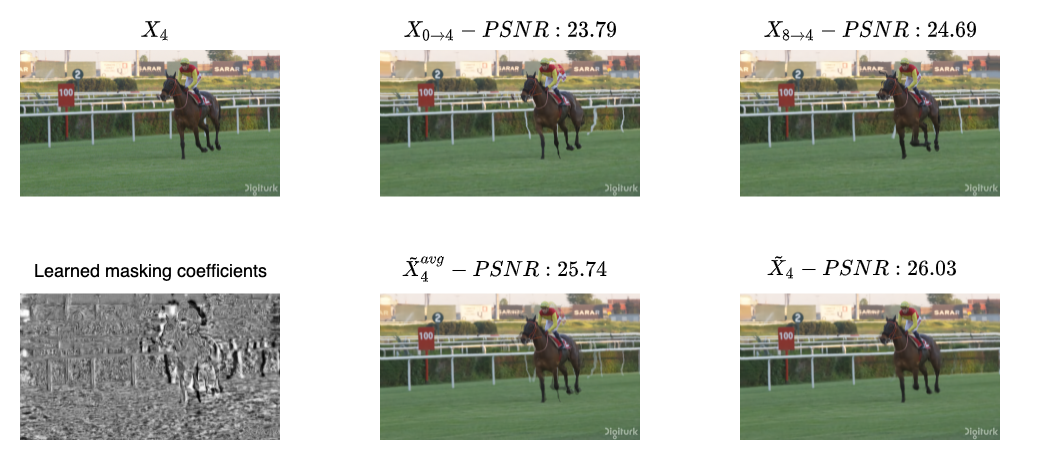} \vspace{-10pt}
    \caption{Effect of learned motion compensation for a hierarchy level $1$ frame in the first GOP of the \textit{Jockey} sequence.}
    \label{fig:jockey_motion_compensation}
\end{figure}

We demonstrate the improvement that comes from a learned fusion mask by means of an example in Figure~\ref{fig:jockey_motion_compensation} with the sequence \textit{Jockey}. As an example case, we take hierarchy level~1 prediction, where we predict the middle frame 4 given frames 0 and 8. 
The past reference frame  warped towards the current frame $X_{0 \rightarrow 4}$ yields $23.79$ dB PSNR, while the future reference frame warped towards the current frame $X_{8 \rightarrow 4}$ gives $24.69$ dB. A simple fusion of these two predictions by averaging, i.e., $\tilde{X}^{avg}_{4} = 0.5 \times (X_{0 \rightarrow 4} + X_{8 \rightarrow 4})$, corresponds to masking with constant coefficients, which are $0.5$ for all pixels. It can be seen from Figure~\ref{fig:jockey_motion_compensation} that even this simple fusion brings more than $1$ dB PSNR gain when compared to individual motion compensated frames. Moreover, when the averaging operation is replaced by the learned pixel-adaptive coefficients, an additional gain is observed, where the PSNR increases from $25.74$ dB to $26.03$ dB. When Figure~\ref{fig:jockey_motion_compensation} is carefully inspected visually, we can see that both motion compensated frames $X_{0 \rightarrow 4}$ and $X_{8 \rightarrow 4}$ contain warping artefacts near the edges of the horse and poles in the background. When these two frames are averaged and $\tilde{X}^{avg}_{4}$ is formed, the overlaid frame still has some visible artefacts near the poles. When learned coefficients are used for fusion, these artefacts are blurred out, yielding a smoother motion compensated frame. 

\begin{figure}[!t]
	\centering
 \includegraphics[width=0.85\linewidth]{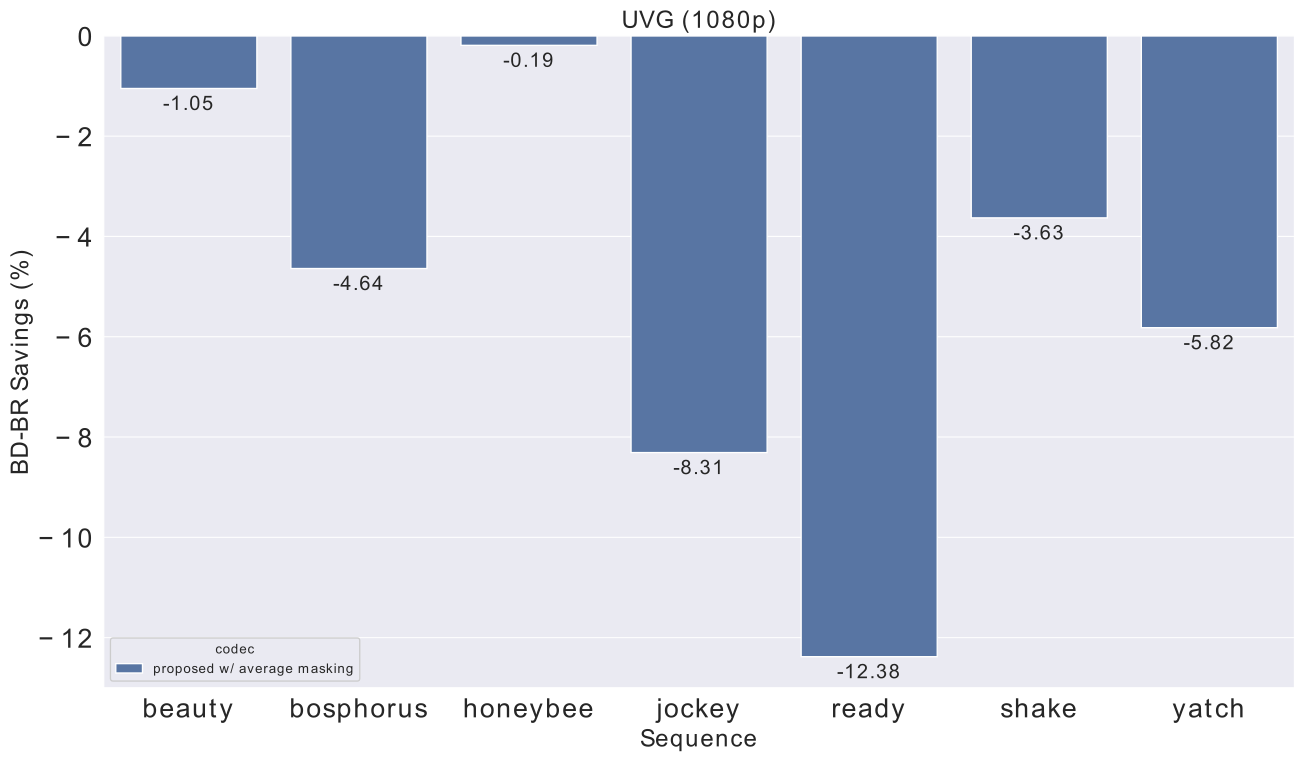}\vspace{-10pt}
    \caption{BD-BR reduction of proposed method relative to average masking}
    \label{fig:proposed_bdbr_sequence_masking}
\end{figure}

We show per sequence BD-BR reductions due to learned masking in bi-directional motion compensation in Figure~\ref{fig:proposed_bdbr_sequence_masking}.  It can be seen that learned masking results in improved performance in all sequences. The improvements are significant especially for high motion sequences with occlusion regions, such as \textit{Jockey} and \textit{Ready}.

\begin{figure}[!t]
	\centering
 \includegraphics[width=1\linewidth]{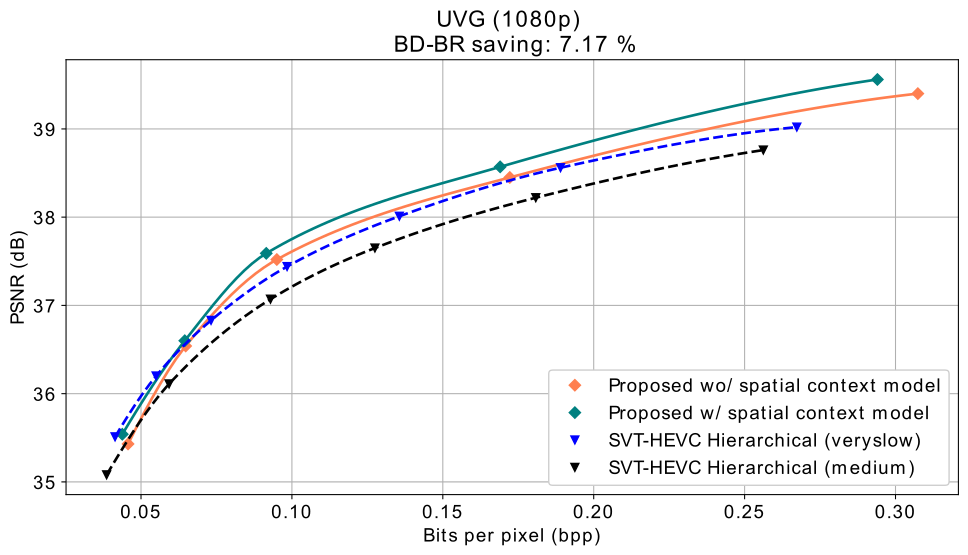} \vspace{-20pt}
    \caption{Effect of backward-adaptive context model on the R-D curve}
    \label{fig:proposed_rd_curve_context}  
\end{figure}

\subsubsection{The Effect of the Backward-Adaptive Context Model}
\begin{figure}[t!]
	\centering
 \includegraphics[width=0.8\linewidth]{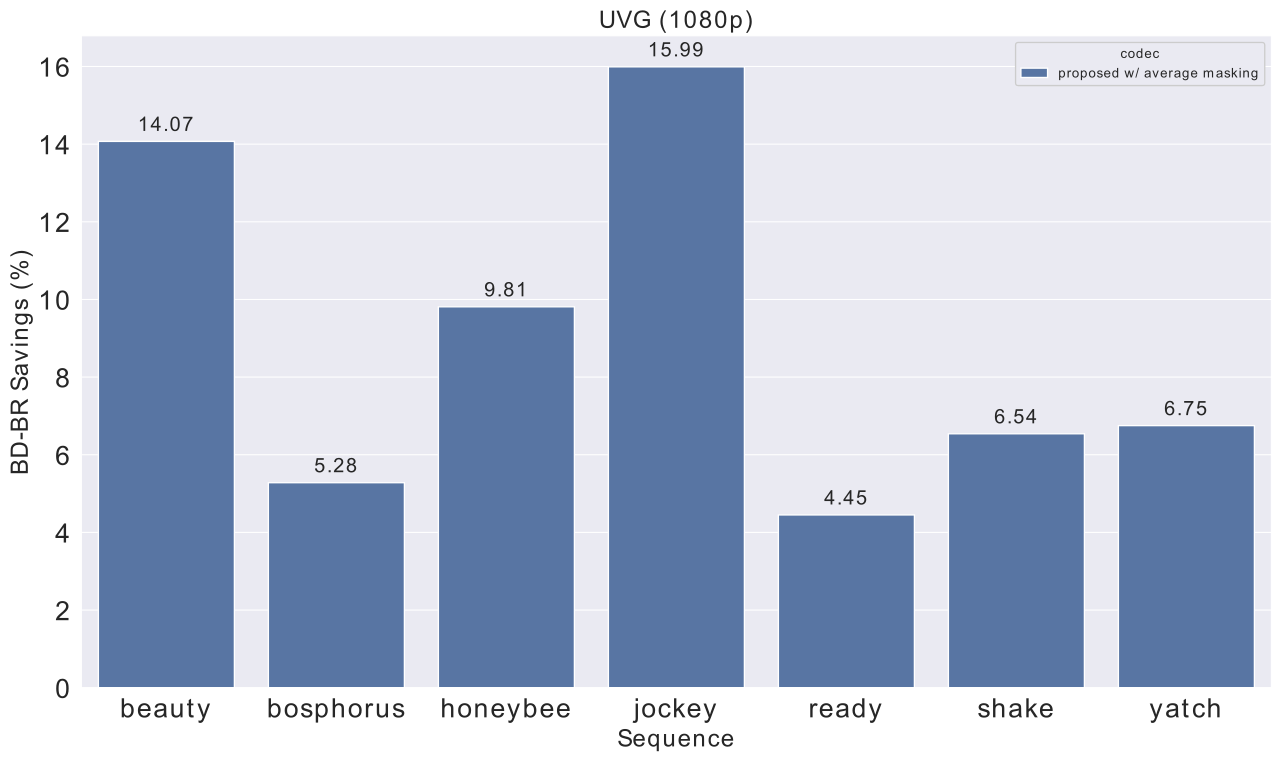} \vspace{-8pt}
    \caption{BD-BR values without the context model vs. with the context model.}
    \label{fig:proposed_bdbr_sequence_context}
\end{figure}

The inclusion of backward-adaptive context model actually improves the rate distortion performance of our codec as shown in Figure~\ref{fig:proposed_rd_curve_context}.  
In particular, when context modeling is omitted, we see $7.17 \%$ average BD-BR increase on the UVG dataset.
We can also see from Figure~\ref{fig:proposed_bdbr_sequence_context} that the context model improves the coding performance in all sequences. 

\begin{figure}[t!]
	\centering
 \includegraphics[width=0.75\linewidth]{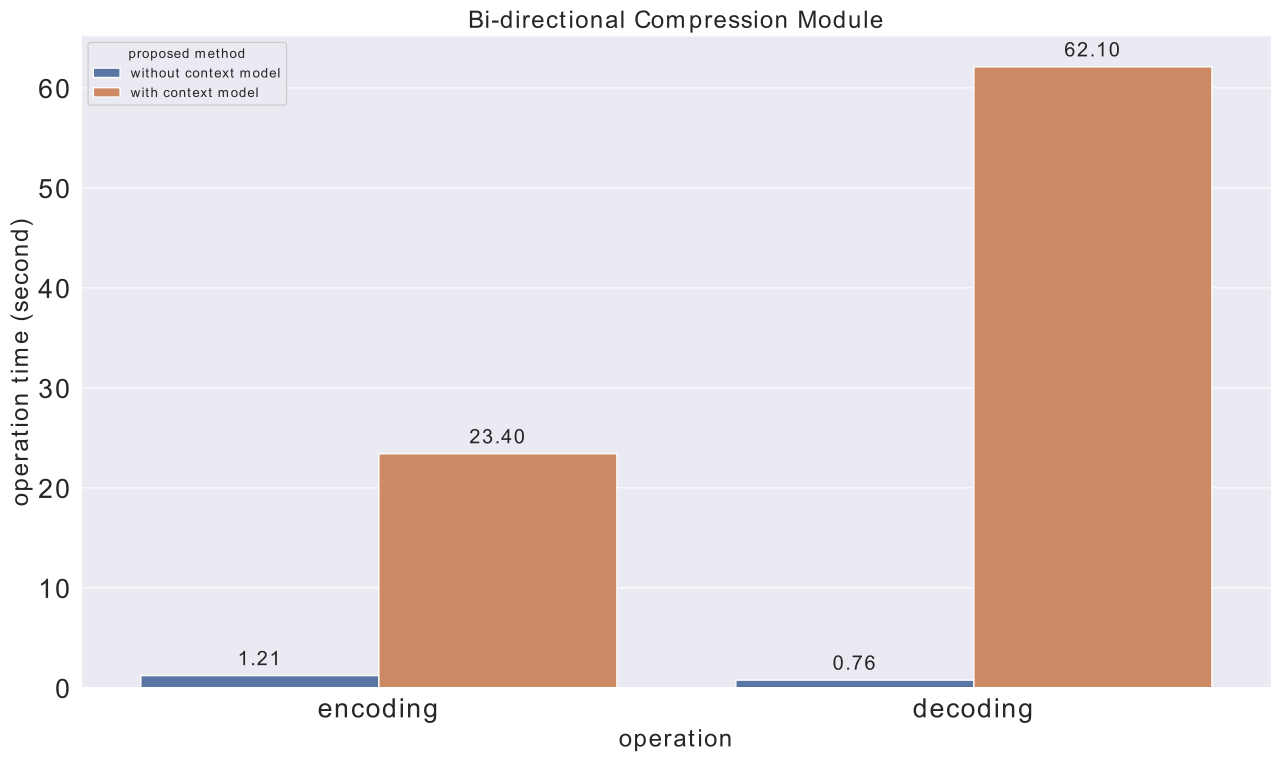}\vspace{-10pt}
    \caption{Average encoding and decoding times of the proposed codec with and without the context model for bi-directional compression of a single frame.}
    \label{fig:proposed_operation_time_context}
\end{figure}

However, this comes at the cost of significant computational complexity and unacceptable decoding times. Encoding and decoding times of the proposed codec with and without context modeling on a single RTX 2080 GPU card are shown in Figure~\ref{fig:proposed_operation_time_context}. Despite its effectiveness, we decided not to include the context model in our default codec configuration since it is not practical to employ a spatial context model in video compression due to its decoding complexity.

\subsubsection{Frame-by-Frame Rate and Distortion Performance}
\begin{figure}[!t]
	\centering
    \includegraphics[width=0.98\linewidth]{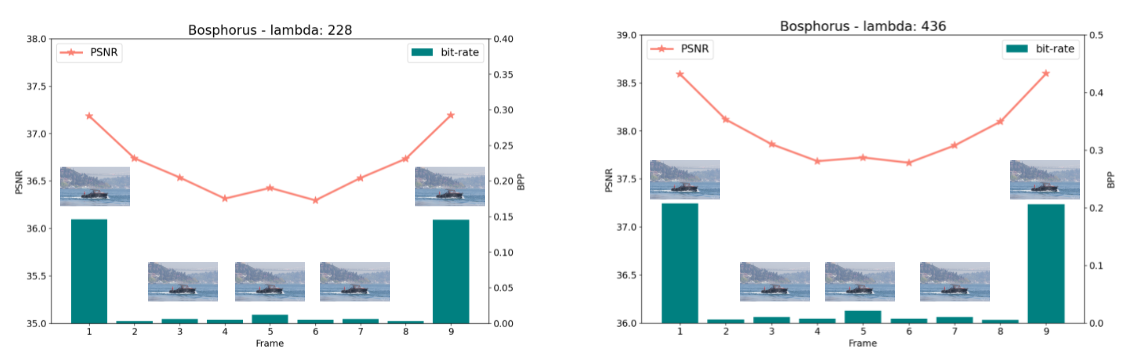} \vspace{-8pt} \\
	(a) \vspace{5pt}\\
	\includegraphics[width=0.98\linewidth]{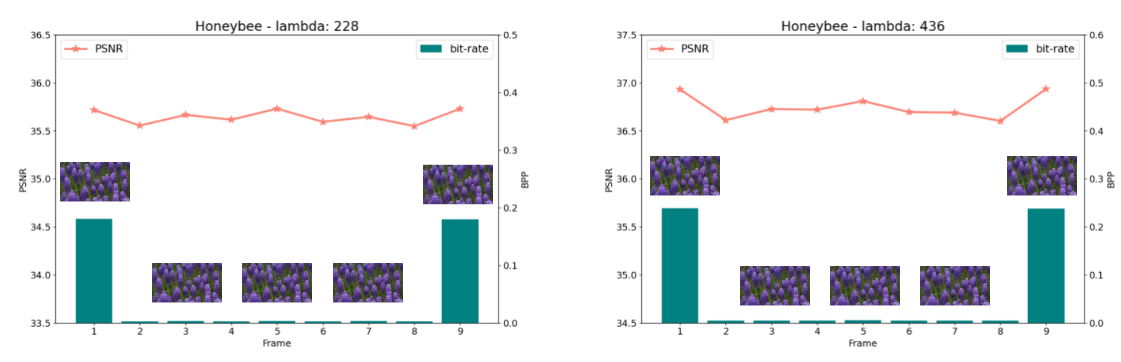} \vspace{-8pt} \\
	(b) \vspace{-8pt}
    \caption{Rate-distortion performance of the proposed model for each frame in the first GOP of the sequences (a): \textit{Bosphorus} (b): \textit{Honeybee}.}
    \label{fig:bosp_honey_gop}
\end{figure}

We can analyze the distribution of rate in bits per pixel (bpp) per frame and PSNR per frame within a GOP, through the~frame-by-frame performance plots shown in Figure~\ref{fig:bosp_honey_gop} for the first GOPs of \textit{Bosphorus} and \textit{Honeybee} sequences. The frame-by-frame performance plots are depicted for two different $\lambda$ values corresponding to different R-D points.

The results show that the PSNR values are more uniform across the frames in \textit{Honeybee} compared to \textit{Bosphorus}. This is because \textit{Honeybee} contains small motion confined to a small area making it easier to compress for all levels of the hierarchy. As expected, key frames contribute most to the overall bit-rate while $B$ frames consume a negligible amount of bits.
On the other hand, the \textit{Bosphorus} sequence contains global camera motion. Hence, the bit-rate gradually increases as the distance to the reference frame increases, since the dynamic range of flow vectors increases. It can be seen that the bit-rate of frame $5$ is greater than that of other $B$ frames. In addition, the PSNR of frames varies and all $B$ frames have a smaller PSNR value compared to $I$ frames. We may balance the PSNR values across all frames by compressing $I$ frames with a higher compression ratio (by changing $\lambda$ values). However we found that reducing the quality of $I$ frames drastically affects the quality of $B$ frames, thus reducing the overall R-D performance. Thus, an unbalanced distribution of PSNR in general yields the best overall R-D performance.

\section{Conclusion}
\label{sec:conclude}
We propose a learned hierarchical bi-directional video compression (LHBDC) framework for end-to-end R-D optimization of hierarchical bi-directional video compression.
Experimental results demonstrate that LHBDC provides significantly improved R-D performance compared to all previous learned codecs as well as the veryslow preset of the traditional  x265 and medium preset of hierarchical SVT-HEVC encoders by comfortable margins. We also outperform the HM~16.23 reference software in the MS-SSIM distortion measure.
Ablation studies show that the proposed innovations, such as motion field subsampling, temporal motion prediction, and learned motion compensation mask, each help improving compression efficiency by  $8.19 \%$, $8.02 \%$, and $6.73 \%$ BD-BR reductions, respectively. Overall, the proposed LHBDC provides near $13 \%$ average BD-BR reduction over hierarchical coding with the medium preset of SVT-HEVC on the UVG testset.

We also show that we give up $7.17 \%$ BD-BR reduction by not including backward-adaptive auto-regressive context modeling. However, backward-adaptive spatial context model is not suitable for use in practical video codecs since it leads to a formidable arithmetic decoder run-time.

Although it has been shown that scale-space motion modeling leads to performance improvements in sequential motion compensation, we show that the proposed learned masking for bi-directional motion compensation is very effective in avoiding artifacts near occlusion regions; hence, eliminating the need for scale-space motion modeling.

Even though we report superior results, LHBDC has some inherent handicaps compared to standards-based codecs: i) we work with RGB 4:4:4 video as opposed to YCrCb 4:2:0 video, ii)~we use an intra-coded key frame at every 8 frames, and iii)~we use no post-processing filtering. Hence, there is still room for further improvement, which is left for future research.

While learned video decoders are currently slower than traditional decoders in todays general purpose hardware, including GPUs, it is expected that special purpose neural decoding hardware that will be available in the near future will overcome this problem.

\appendices

\section{Low-latency compression with x265 codec}
\label{appendixA}
We used the x265 encoder embedded in ffmpeg to generate compressed video sequences for low-latency compression which does not utilize $B$ frames. The command line is: 

\textit{ffmpeg -pix\_fmt yuv420p -s 1920$\times$1080 -i video.yuv -c:v libx265 -preset veryslow -tune zerolatency -x265-params "crf=Q:keyint=10" video.mkv}

In the command above, \textit{crf:Q} represents constant rate factor and affects the~quality of generated videos. A higher \textit{Q} means a higher compression rate and a lower quality.

\section{Hierarchical compression with SVT-HEVC codec}
\label{appendixB}
SVT-HEVC encoder is optimized for the classical hierarchical B-frame compression (see \url{https://github.com/OpenVisualCloud/SVT-HEVC}).
The command line is:

\textit{ffmpeg -pix\_fmt yuv420p -s 1920$\times$1080 -i video.yuv -c:v libsvt\_hevc -preset 3 -g 8 -hielevel 2 -qp QP video.mkv}

\noindent where \textit{-preset} option determines the encoding speed and is set to $3, 6$ in order to perform coding at veryslow and medium levels respectively. \textit{-g} specifies the GOP size set to $8$ for comparison with our proposed codec. \textit{-hielevel} enables multiple hierarchical levels in SVT-HEVC. It is set to~$2$ indicating a $3$ level hierarchy for the selected GOP size. \textit{QP} represents the quantization parameter controlling rate-distortion trade-off.

\section{Hierarchical Compression with HM-16.23 Codec}
\label{appendixC}
We employed the random access configuration of HEVC reference software HM16.23 given by \url{https://vcgit.hhi.fraunhofer.de/jvet/HM/-/blob/HM-16.23/cfg/encoder_randomaccess_main.cfg} with only modifying \textit{GOPSize} and \textit{IntraPeriod} equal to $8$ and setting the reference frames for hierarchical B-frame compression accordingly. The same configuration is used for 7 QP values: 19, 21, 23, 25, 27, 29, 32 to plot the R-D curve.







%


\printbibliography


  

%

\vfill
\begin{IEEEbiography}[{\includegraphics[width=1in,height=1.25in,clip,keepaspectratio]{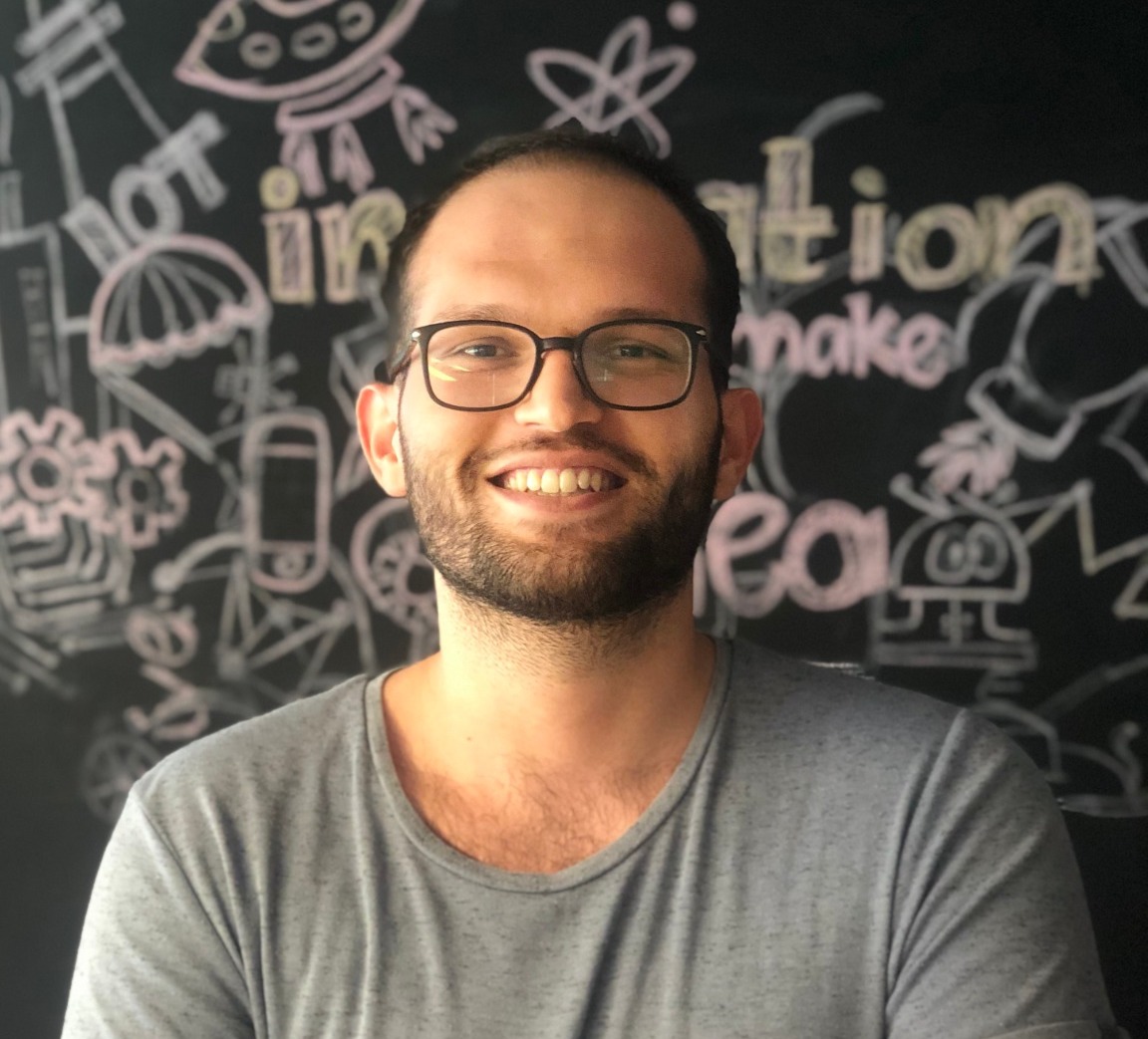}}]
{M. Akın Yılmaz} (S'19-M'21) received M.S. degree in Electrical and Electronics Engineering Koc University in 2021. His research interests focus on learned image and video compression.
\end{IEEEbiography}

\vfill
\begin{IEEEbiography}[{\includegraphics[width=1in,height=1.25in,clip,keepaspectratio]{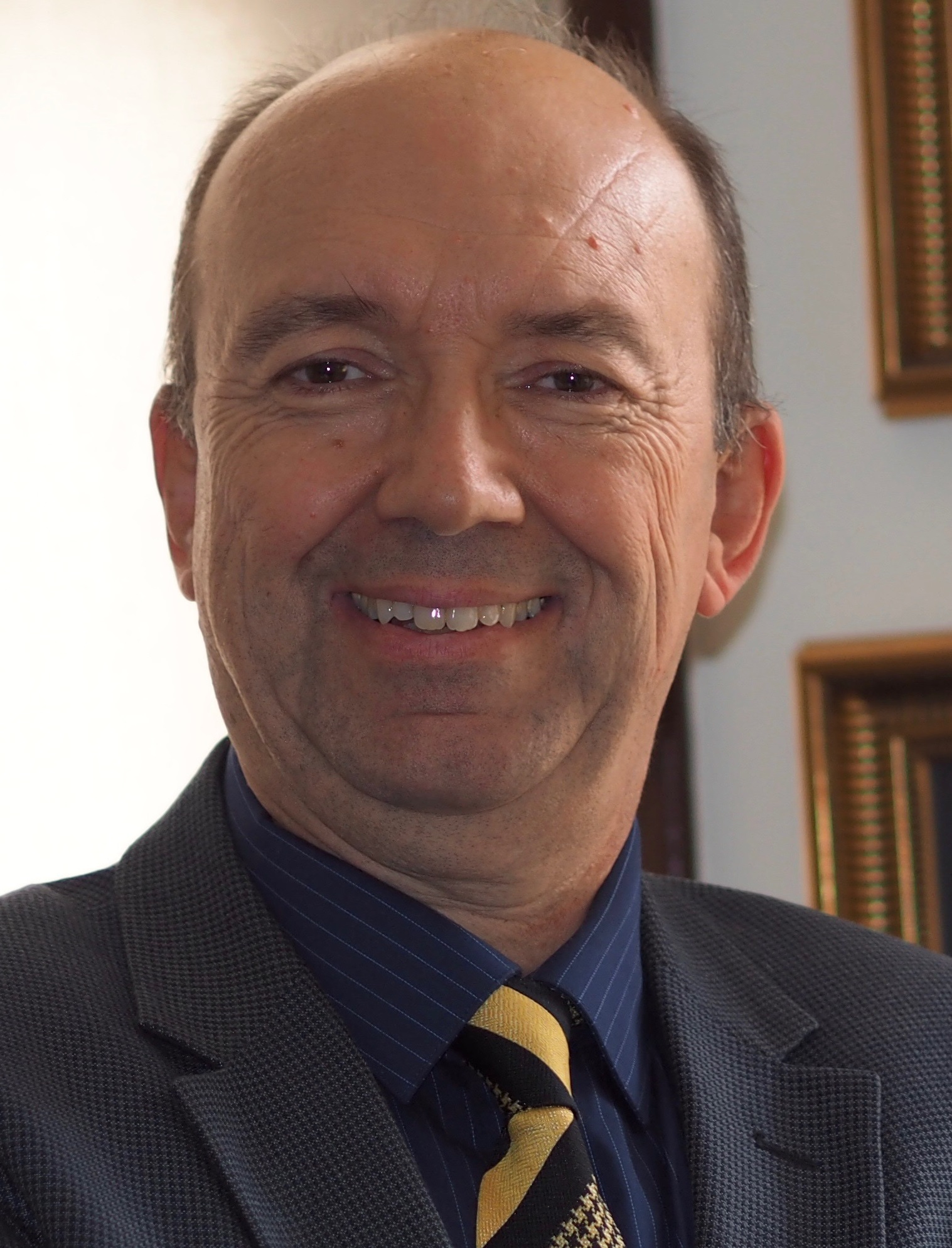}}]{A. Murat Tekalp} (S'80-M'84-SM'91-F'03) received Ph.D. degree in Electrical, Computer, and Systems Engineering from Rensselaer Polytechnic Institute (RPI), Troy, New York, in 1984, He was with Eastman Kodak Company, Rochester, New York, from 1984 to 1987, and with the University of Rochester, Rochester, New York, from 1987 to
2005, where he was promoted to Distinguished University Professor. He is currently Professor at Koc University, Istanbul, Turkey. He served as Dean of Engineering between 2010-2013. His research interests
are in digital image and video processing, including video compression
and streaming, video networking, multi-view and 3D video processing, and
deep learning for image/video processing and compression.

He has been elected a member of Turkish Academy of Sciences and
Academia Europaea. He served as an Associate Editor for the IEEE Trans. on Signal Processing (1990-1992) and IEEE Trans. on Image Processing (1994-1996). He was the Editor-in-Chief of the EURASIP journal Signal Processing: Image Communication published by Elsevier between 1999-2010. He was on the Editorial Board of the IEEE Signal Processing Magazine (2007-2010) and the Proceedings of the IEEE (2014-2020). He chaired the IEEE Signal Processing Society Technical Committee on Image and Multidimensional Signal Processing (Jan. 1996 - Dec. 1997). He was appointed as the General Chair of IEEE International Conference on Image Processing (ICIP) at Rochester, NY in 2002. He served in the European Research Council (ERC) Advanced Grant Panels (2009-2015). He was the Technical Program Co-Chair for IEEE ICIP~2020. He is currently in the Editorial Board of Wiley-IEEE Press. Dr. Tekalp has authored the Prentice Hall book Digital Video Processing~(1995), a completely rewritten second edition of which is published by Pearson in 2015. 
\end{IEEEbiography}




\end{document}